\documentclass{amsart}
\usepackage{amsfonts}

\usepackage{graphicx}
\usepackage{amscd}

\newtheorem{theorem}{Theorem}
\theoremstyle{plain}

\newtheorem{lemma}{Lemma}

\numberwithin{equation}{section}
\numberwithin{theorem}{section}
\numberwithin{lemma}{section}
\numberwithin{proposition}{section}
\numberwithin{corollary}{section}

\input{tcilatex}

\begin{document}
\title[$p-$adic pseudo-differential operators]{Fundamental solutions of pseudo-differential operators over $p-$adic fields}
\author{W. A. Zuniga-Galindo}
\address{Department of Mathematics and Computer Science, Barry University, 11300 N.E.
Second Avenue, Miami Shores, Florida 33161, USA}
\email{wzuniga@mail.barry.edu}

\begin{abstract}
We show the existence of fundamental \ solutions for $p-$adic
pseudo-differential operators with polynomial symbols.
\end{abstract}

\maketitle

\section{\protect\bigskip Introduction}

Let $K$ be a $p-$adic field, i.e. a finite extension of $\mathbb{Q}_{p}$ the
field of $p-$adic numbers. Let $R_{K}$ be the valuation ring of $K$, $P_{K}$
the maximal ideal \ of \ $R_{K}$, and \ $\overline{K}=R_{K}/$ $P_{K}$ the
residue field \ of $K$. The cardinality of $\overline{K}$ is denoted by $q$.
For $z\in K$, $v(z)\in \mathbb{Z}\cup \left\{ +\infty \right\} $ denotes the
valuation of $z$, $\left| z\right| _{K}=q^{-v(z)}$ and $ac(z)=z\pi ^{-v(z)}$
where \ $\pi $ is a fixed uniformizing parameter for \ $R_{K}$. Let $\Psi $
denote an \ additive character of $K$ trivial on $R_{K}$ but not on $%
P_{K}^{-1}$.

A function $\Phi :K^{n}\rightarrow \mathbb{C}$ \ \ is called a
Schwartz-Bruhat function if it is \ locally constant with compact support.
We denote by $\mathcal{S}(K^{n})$ the $\mathbb{C}$-vector space of
Schwartz-Bruhat functions over $K^{n}$.The dual \ space $\mathcal{S}^{\prime
}(K^{n})$ is the space of distributions over $K^{n}$. Let \ $f=f\left(
x\right) \in K\left[ x\right] $, $x=\left( x_{1},..,x_{n}\right) $, be a
non-zero polynomial, and $\beta $ a \ complex number satisfying $\func{Re}%
(\beta )>0$. If $x=\left( x_{1},...,x_{n}\right) ,$ $y=\left(
y_{1},...,y_{n}\right) \in K^{n}$, we set $\left[ x,y\right] :=$ $%
\sum_{i=1}^{n}x_{i}y_{i}$.

A $p$-adic pseudo-differential operator $f(\partial ,\beta )$, with symbol $%
\left| f\right| _{K}^{\beta }$, is an operator of the form

\begin{equation}
\begin{array}{cccc}
f(\partial ,\beta ): & \mathcal{S}(K^{n}) & \rightarrow & \mathcal{S}(K^{n})
\\ 
& \Phi & \rightarrow & \mathcal{F}^{-1}\left( \left| f\right| _{K}^{\beta }%
\mathcal{F}\left( \Phi \right) \right) ,
\end{array}
\end{equation}
where \ 
\begin{equation}
\begin{array}{cccc}
\mathcal{F}: & \mathcal{S}(K^{n}) & \rightarrow & \mathcal{S}(K^{n}) \\ 
& \Phi & \rightarrow & \int\limits_{K^{n}}\Psi \left( -\left[ x,y\right]
\right) \Phi \left( x\right) dx
\end{array}
\end{equation}
\ is the Fourier transform. The operator \ $f(\partial ,\beta )$\ has
self-adjoint extension with dense domain in $L^{2}\left( K^{n}\right) $. \
We associate to $f(\partial ,\beta )$ the following \ $p$-adic
pseudo-differential equation: 
\begin{equation}
f(\partial ,\beta )u=g\text{, \ }g\in \mathcal{S}(K^{n}).  \label{for2}
\end{equation}

A fundamental solution for (\ref{for2}) is a distribution $E$ such that \ $%
u=E\ast g$ is a solution.

The main result of this paper is the following.

\begin{theorem}
\label{th1}Every $p$-adic pseudo-differential equation $f(\partial ,\beta
)u=g,$ with $f\left( x\right) $ $\in $ $K\left[ x_{1},..,x_{n}\right]
\setminus K$, \ $g\in \mathcal{S}(K^{n})$, \ and $\beta $ $\in \mathbb{C}$, $%
\func{Re}(\beta )>0$, has \ a fundamental solution $E\in \mathcal{S}^{\prime
}(K^{n})$.
\end{theorem}

The \ $p-$adic pseudo-differential operators \ occur naturally in $p-$adic
quantum field theory \ \cite{VVZ}, \cite{Koch1}. Vladimirov showed the
existence of a fundamental solutions for symbols of the form $\left| \xi
\right| _{K}^{\alpha }$, $\alpha >0$ \cite{V}, \ \cite{VVZ}. In \cite{Koch2}%
, \cite{Koch1}\ Kochubei \ showed explicitly the existence of fundamental
solutions for operators \ with symbols \ of the form $\left| f\left( \xi
_{1},..,\xi _{n}\right) \right| _{K}^{\alpha }$, $\alpha >0$, \ where \ $%
f\left( \xi _{1},..,\xi _{n}\right) $ is a quadratic form satisfying $%
f\left( \xi _{1},..,\xi _{n}\right) \neq 0$ \ if $\left| \xi _{1}\right|
_{K}^{{}}+..+\left| \xi _{n}\right| _{K}^{{}}\neq 0$. In \cite{Khr}
Khrennikov considered spaces of functions and distributions defined outside
\ the singularities \ of a symbol, in this situation he showed the existence
of a fundamental solution for a $p-$adic pseudo-differential equation with
symbol $\left| f\right| _{K}\neq 0$. The main result of this paper shows the
existence of fundamental solutions for operators with \ polynomial symbols.
Our proof is based on a solution of the division problem for $p-$adic
distributions. This problem is solved by \ adapting the ideas \ developed by
Atiyah for the archimedean case \cite{A}, and Igusa's theorem on the
meromorphic continuation of local zeta functions \cite{I1}, \cite{I2}. The
connection between local zeta functions (also called Igusa's local zeta
functions) and \ fundamental solutions of $p-$adic pseudo-differential
operators has been explicitly showed in particular cases by Jang and Sato\ 
\cite{J},\ \cite{S}. \ In \cite{S} \ Sato studies the \ asymptotics of \ the
Green function $G$ of the following pseudo-differential equation 
\begin{equation}
\left( f\left( \partial ,1\right) +m^{2}\right) u=g\text{, \ }m>0.
\end{equation}

The main result in \cite[ theorem 2.3]{S} describes the asymptotics of $G(x)$
when the polynomial $f$ is a relative invariant of some prehomogeneous
vector spaces (see e.g. \cite[Chapter 6]{I1}). The key step\ is to establish
a connection \ between the \ Green function $G(x)$ and the $\ $\ local zeta
function attached to $f$.

All the \ above mentioned results suggest a deep connection between Igusa's
work on local zeta functions (see e.g. \cite{I1}) and $p-$adic
pseudo-differential equations.

\section{Local zeta functions and division of distributions}

The local zeta function associated to $f$\ is the distribution 
\begin{equation}
\left\langle \left| f\right| _{K}^{s},\Phi \right\rangle
=\int\limits_{K^{n}\setminus f^{-1}(0)}\Phi \left( x\right) \left| f\left(
x\right) \right| _{K}^{s}dx,  \label{for3}
\end{equation}
where \ $\Phi \in \mathcal{S}(K^{n})$, $s\in \mathbb{C}$, $\func{Re}(s)>0$,
and $dx$\ is the Haar of $K^{n}$ normalized so that $vol\left(
R_{K}^{n}\right) =1$. The local zeta functions were introduced by Weil \cite
{W} and their basic properties for general $f$ were first studied by Igusa 
\cite{I1}, \cite{I2}. A central result in the theory of local zeta functions
is the following.

\begin{theorem}[{Igusa, \protect\cite[Theorem 8.2.1]{I1}}]
\label{th2}The \ distribution $\left| f\right| _{K}^{s}$ admits a
meromorphic continuation \ to the complex plane such that \ $\left\langle
\left| f\right| _{K}^{s},\Phi \right\rangle $ is a rational function of $%
q^{-s}$ for each $\Phi \in \mathcal{S}(K^{n})$. In addition the real parts
of the poles \ of \ $\left| f\right| _{K}^{s}$ \ are negative rational
numbers.
\end{theorem}

The archimedean counterpart of the previous theorem \ was obtained \ jointly
\ by Bernstein and \ Gelfand \cite{B1}, independently by \ Atiyah \cite{A}.
The following lemma is a consequence of the previous theorem.

\begin{lemma}
\label{lem1}Let $f\left( x\right) $ $\in K\left[ x_{1},..,x_{n}\right] $ be
a non-constant polynomial, \ and $\beta $ a \ complex number satisfying $%
\func{Re}(\beta )>0$. Then \ there exists a distribution $T\in \mathcal{S}%
^{\prime }(K^{n})$ satisfying \ $\left| f\right| _{K}^{\beta }T=1$.
\end{lemma}

\begin{proof}
By theorem \ref{th2} $\left| f\right| _{K}^{s}$ \ has a meromorphic
continuation to $\mathbb{C}$ \ such that \ $\left\langle \left| f\right|
_{K}^{s},\Phi \right\rangle $ is a rational function of $q^{-s}$ for each $%
\Phi \in \mathcal{S}(K^{n})$. Let 
\begin{equation}
\left| f\right| _{K}^{s}=\sum\limits_{m\in \mathbb{Z}}^{{}}c_{m}\left(
s+\beta \right) ^{m}  \label{for4}
\end{equation}
be the Laurent expansion at $-\beta $ with $c_{m}\in \mathcal{S}^{\prime
}(K^{n})$ for all $m$. Since \ the real parts of the poles of $\left|
f\right| _{K}^{s}$ are negative rational numbers by theorem \ref{th2}, it
holds that $\left| f\right| _{K}^{s+\beta }=\left| f\right| _{K}^{\beta }\
\left| f\right| _{K}^{s}$\ \ is holomorphic at $s=-\beta $. Therefore $%
\left| f\right| _{K}^{\beta }c_{m}=0$ for all $m<0$ and 
\begin{equation}
\left| f\right| _{K}^{s+\beta }=c_{0}\left| f\right| _{K}^{\beta
}+\sum\limits_{m=1}^{\infty }c_{m}\left| f\right| _{K}^{\beta }\left(
s+\beta \right) ^{m}.  \label{for5}
\end{equation}
By using the Lebesgue lemma and (\ref{for5}) it holds that 
\begin{eqnarray}
\ \lim_{s\rightarrow -\beta }\left\langle \left| f\right| _{K}^{s+\beta
},\Phi \right\rangle  &=&\int\limits_{K^{n}}\Phi \left( x\right)
dx=\left\langle 1,\Phi \right\rangle   \notag \\
&=&c_{0}\left| f\right| _{K}^{\beta }.  \label{for6}
\end{eqnarray}
\ \ Therefore we can take $T=c_{0}$.
\end{proof}

\bigskip

If $T\in \mathcal{S}^{\prime }\left( K^{n}\right) $ we denote by \ $\mathcal{%
F}T\in \mathcal{S}^{\prime }\left( K^{n}\right) $ the Fourier \ transform of
the distribution $S$, i.e. $\left\langle \mathcal{F}T,\Phi \right\rangle $ $%
=\left\langle S,\mathcal{F}\left( \Phi \right) \right\rangle $, $\Phi \in 
\mathcal{S}\left( K^{n}\right) $.

\section{Proof of the main result}

By lemma \ref{lem1} there exists a $T\in \mathcal{S}^{\prime }\left(
K^{n}\right) $ such that $\left| f\right| _{K}^{\beta }T=1$. We set $E=%
\mathcal{F}^{-1}T\in \mathcal{S}^{\prime }\left( K^{n}\right) $ and assert
that $E$ is \ a fundamental solution for (\ref{for2}). This last statement
is equivalent to assert that $\mathcal{F}\left( \Phi \right) =\left( 
\mathcal{F}E\right) \mathcal{F}\left( g\right) $ satisfies $\left| f\right|
_{K}^{\beta }\mathcal{F}\left( \Phi \right) =\mathcal{F}\left( g\right) $.
Since $\left| f\right| _{K}^{\beta }\mathcal{F}\left( \Phi \right) =\left|
f\right| _{K}^{\beta }\left( \mathcal{F}E\right) \mathcal{F}\left( g\right) $
$=\left| f\right| _{K}^{\beta }T$ $\mathcal{F}\left( g\right) =\mathcal{F}%
\left( g\right) $, we have that $E$ is \ a fundamental solution for (\ref
{for2}).

\section{\protect\bigskip Operators with twisted \ symbols}

Let $\chi :R_{K}^{\times }\rightarrow \mathbb{C}$ be a non-trivial
multiplicative character, i.e. \ a homomorphism with finite image, where $%
R_{K}^{\times }$ is the group\ of units of $R_{K}$. We put formally $\chi
\left( 0\right) =0$. If $f\left( x\right) \in $ $K\left[ x_{1},..,x_{n}%
\right] \setminus K$, we say that $\chi \left( ac(f)\right) \left| f\right|
_{K}^{\beta }$, with $\beta \in \mathbb{C}$, $\func{Re}(\beta )>0$, is a 
\textit{twisted symbol}, and \ call the \ pseudo-differential operator \ 
\begin{equation}
\Phi \rightarrow f(\partial ,\beta ,\chi )\Phi =\mathcal{F}^{-1}\left( \chi
\left( ac(f)\right) \left| f\right| _{K}^{\beta }\mathcal{F}\left( \Phi
\right) \right) ,\Phi \in \mathcal{S}(K^{n}),
\end{equation}
a \textit{twisted operator}. Since the distribution $\chi \left(
ac(f)\right) \left| f\right| _{K}^{\beta }$\ satisfies all the properties
stated in theorem \ref{th2} (cf. \cite[Theorem 8.2.1]{I1}), theorem \ref{th1}
generalizes \ literally \ to the case of twisted operators. In \cite[chapter
2]{Koch1}\ Kochubei showed explicitly the existence of fundamental solutions
for \ twisted operators in some particular cases.

\end{document}